\documentstyle[preprint,aps]{revtex}
\begin{document}

\draft
\title{
\begin{flushright}
{\rm Oslo-TP-4-94,
USITP-94-5\\
gr-qc/9403060}
\end{flushright}
Scattering of Fermions off Dilaton Black Holes}

\author{Bj\o rn Jensen\footnote{Electronic address: BJensen@boson.uio.no}}
\address{
Institute of Physics, University of Oslo,
P.O. Box 1048, N-0316 Blindern, Oslo 3, Norway}
\author{Ulf Lindstr\"{o}m\footnote{Electronic address: UL@vana.physto.se and
ulfl@boson.uio.no}}
\address{ITP, University of Stockholm,
Box 6730, S-113 85, Stockholm, Sweden }
\date{\today}
\bibliographystyle{unsrt}
\maketitle

\begin{abstract}
We discuss how various properties of dilaton black holes depend
on the dilaton coupling constant $a$. In particular
we investigate the $a$-dependence of certain mass parameters both
outside and in the extremal limit and discuss their
relation to thermodynamical quantities. To further illuminate
the role of the coupling constant $a$ we look at a massless point particle
in a dilaton black hole geometry as well as the scattering
of (neutral) fermions. In this latter case we find that the
scattering potential vanishes for the zero angular momentum
mode which seems to indicate a catastrophic deradiation when
$a>1$.
\end{abstract}
\vspace{1cm}
Dilaton black holes are studied because they arise as solutions to the
low-energy string effective field theory \cite{Witten}, and because they shed
new light on the old question of black holes as elementary particles
\cite{Holzhey}.
The dilaton black holes are solutions to field equations that describe a scalar
(dilaton) field and an electromagnetic field coupled to Einstein gravity. It
has proven useful to let the dilaton coupling depend on a non-negative
parameter $a$. The value $a=0$ corresponds to the usual Einstein
theory whereas
 the coupling $a=1$ is the value dictated by string theory. Furthermore, the
value $a=\sqrt 3$
seems to be special. For this value of $a$ rotating solutions \cite{Frolov} as
well
as a string black hole \cite{Duff} have been found. It also arises in
dimensional
reduction of $d=5$ supergravity studied in the context of supersymmetric
solitons
 \cite{Gibbons2}.
In general many features of the solution (but not all) depend on the value of
$a$. The present work corroborates this picture.\\
\smallskip
The main problem treated in this note is the scattering of spinning particles
off of
 dilaton black holes. We derive the corresponding potential and discuss its
behaviour as a function of $a$. In particular we find that it generally
diverges at the horizon in the extremal limit for $a>1$. For $a=\sqrt 3$ there
is an additional pole, interestingly enough, although the relation to the
abovementioned rotating solution, string solution
or soliton solutions is unclear.\\
\smallskip
This note also contains a derivation and discussion of certain mass-parameters:
the ADM-mass, the Tolman mass and the mass of the black hole. Although we
relate the usual definition of the temperature and entropy of the black hole to
its
ADM mass,
 we use the vanishing of the black hole mass as the main indication for the
vanishing of the black hole in the extremal limit. This is because it has been
argued that the thermo-dynamical description breaks down in that limit
\cite{Preskill,Holzhey}.
\newpage
The action we consider is
\begin{equation}
S=\int d^4x\sqrt{-g}(-R+2(\nabla\Psi )^2+e^{-2a\Psi}F^2)\,
\end{equation}
where
$R$ is the Ricci curvature scalar, $\Psi$ the dilaton field, $F^2$ the
square of the
electromagnetic field strength and $a$ a
non-negative real parameter.
The corresponding field equations are
\begin{eqnarray}
&\nabla&_\mu (e^{-2a\Psi}F^{\mu\nu})=0\, ,\\
&\nabla&^2\Psi +\frac{a}{2}e^{-2a\Psi}F^2=0\, ,\\
&R&_{\mu\nu}=2\nabla_\mu\Psi\nabla_\nu\Psi+2e^{-2a\Psi}F_\mu\, _\rho F_\nu\,
^\rho -\frac{1}{2}g_{\mu\nu}e^{-2a\Psi}F^2\, .
\end{eqnarray}
The line element for an electrically charged black hole solution is
\cite{Gibbons,Garfinkle}
\begin{equation}
ds^2=\Delta\sigma^{-2}dt^2-\sigma^2(\Delta^{-1}dr^2+r^2d\Omega^2 _2)\, ,
\end{equation}
with $\Delta$ and $\sigma$ given by
\begin{equation}
\Delta =(1-\frac{r_+}{r})(1-\frac{r_-}{r})\, ,\,
\sigma^2=(1-\frac{r_-}{r})^{\frac{2a^2}{1+a^2}}\, .
\end{equation}
Here
$d\Omega^2 _2$ is the metric in the unit two-sphere $S^2$, $r_+$ and $r_-$ are
implicitly given by
\begin{equation}
r_+r_-=Q^2(1+a^2)\,\,\, ,\,\,\, (2M-r_+)(\frac{1+a^2}{1-a^2})=r_-\, ,
\end{equation}
$M$ represents the metric
mass and $Q$ is the electric charge on the hole.
The electric field is given by
\begin{equation}
\underline{F}=\frac{Q}{r^2}\underline{d}t\wedge\underline{d}r\, ,
\end{equation}
and the dilaton by
\begin{equation}
e^{2a\Psi}=\sigma^2\, .
\end{equation}
The scalar charge $\Sigma$ is defined as $\Psi\sim\Psi_0-\Sigma/r$
for large $r$ which gives
\begin{equation}
r_-=\frac{(1+a^2)\Sigma}{a}\Rightarrow r_+=a\frac{Q^2}{\Sigma}\, .
\end{equation}
The metric mass can thus be written
\begin{equation}
M=\frac{a^2Q^2+(1-a^2)\Sigma^2}{2a\Sigma}\, .
\end{equation}
Some aspects of these solutions can be better understood
when one analytically continues the Schwartzschild coordinates.
To this end we
define an advanced null coordinate $v$ by $v=t+r^*$ where
\begin{equation}
\frac{dr^*}{dr}=\frac{\sigma^2}{\Delta}\, .
\end{equation}
In terms of this coordinate we easily obtain the advanced
Eddington-Finkelstein form of the metric eq.(5)
\begin{equation}
ds^2=\frac{\Delta}{\sigma^2}dv^2-2dvdr-r^2\sigma^2d\Omega_2 ^2\, .
\end{equation}
 It is clear that
this metric in general
is free from singularities when $a\leq 1$.
Off extremality ($r_-\neq r_+$) the analytic continuation
is singular at $r=r_-$ when $a>1$. However this
singularity generally
disappears in the extremal limit, i.e. in this limit
the metric is non-singular all the way
down to the singularity at $r=0$.
We finally give the Kruskal form of the metric. Defining the ($U,V$)
coordinates in the usual way we get
\begin{equation}
ds^2=-32M^3\frac{d}{dr}e^{-\frac{r^*}{2M}}(
dU^2-dV^2)-r^2\sigma^2d\Omega^2_2\, .
\end{equation}
{}From this expression one can easily convince oneself
that  the ($U,V$) part of the metrics of the $a=0$ and $a=1$ solutions
are in fact identical (when $2M$ is replaced by $r_+$ in the
$a=1$ solution). Hence, when $a=1$ the sole effect of the
 singularity at $r=r_-$, (which
coincides with the event-horizon in the extremal limit),
is the vanishing
of the angular part of the metric. The structure of the $a\neq 1$ extensions is
much more complicated and will not be considered here.\\
\smallskip
Now we want to discuss the energy content of the
solutions and to this end we calculate certain mass parameters.
Let $\xi^a$ be a time translation Killing vector field which is timelike near
infinity such that
$\xi^a\xi_a=1$ and which has vanishing norm $\xi^a\xi_a=0$ at the horizon.
Let $S$ denote the region outside the event-horizon. The boundary $\partial S$
of $S$ is taken to be the event-horizon $\partial B$ of the black hole
 and a two-surface $\partial S_\infty$
at infinity.
Due to the asymptotic flatness of the dilaton black hole solutions
it follows that the ADM mass of the black hole
configuration is given by \cite{Bardeen}
\begin{eqnarray}
M^\infty &=&\frac{1}{8\pi}\int_{S}(2T^a\, _b -T\delta^a\,
_b)\xi^bd\Sigma_a+\frac{1}{4\pi}\int_{\partial B}
\xi^{a;b}d\Sigma_{ab}
\equiv M^M+M^H\, .
\end{eqnarray}
The last integral is just the surface gravity $\kappa$ multiplied with the
surface area $A^H$ of
the event-horizon. $M^M$ and $M^H$ are naturally interpreted as the
gravitational
mass of the matter distribution
outside the event-horizon and the gravitational mass of the
black hole, respectively.
 In the following we will sett
$\vec{\xi}=\partial_t$ since this vector always vanishes on the
event-horizon independent of the value of $a$.
It then follows that the first integral $M^M$
is the usual Tolman mass of the matter outside the event-horizon \cite{Tolman}.
{}From the action eq.(1) we derive the energy-momentum tensor
$T_{\mu\nu}$ for the dilaton and the electromagnetic field
\begin{eqnarray}
T^t\, _t&=& \frac{\sigma^2}{\Delta}(\frac{\Delta Q^2}{\sigma^4
r^4}+\frac{\Sigma^2\Delta^2}{\sigma^4r^2(r-r_-)^2})\, ,\\
T^r\, _r&=&
-\frac{\Delta}{\sigma^2}(\frac{\Sigma^2}{r^2(r-r_-)^2}-\frac{Q^2}{\Delta
r^4})\, ,\\
T^\Omega\, _\Omega &=& -\frac{Q^2}{\sigma^2r^4}+\frac{\Delta\Sigma^2}{
\sigma^2r^2(r-r_-)^2}\, .
\end{eqnarray}
 The Tolman mass density is then given by
\begin{eqnarray}
{\cal M}^M&=&T^t\, _t-T^r\, _r-2T^\Omega\, _\Omega =
\frac{2Q^2}{\sigma^2r^4}
\, .
\end{eqnarray}
This expression reduces to the usual one in the Reissner-Nordstr\"{o}m geometry
when $a$ is sett to zero. From the above expression it follows
that ${\cal M}^M$ generally diverges
on $r=r_-$ when
 $a>0$ thus giving
rise to a curvature singularity there.
 However, the total energy outside the
event-horizon is always positive and finite. It is rather surprising that
it is only the electric field that contributes to the gravitational
energy. The dilaton contribution cancels due to the large
 pressures in the angular directions.\\
\smallskip
It is interesting to relate the temperature and the entropy
of the black hole to its ADM mass.
Let $\vec{N}$ denote a second Killing vector field
(normalized to unity $N^aN_a=-1$)
orthogonal to the event-horizon.
The surface gravity $\kappa$ can then be written
$\kappa^2= N_b\xi^a\nabla_a\xi^b$ \cite{Bardeen}.
The area of the event-horizon is $A^H=4\pi r_+^2\sigma^2(r=r_+)$. This
implies that the contribution to the total gravitational mass of the system
from
behind the event-horizon is
\begin{equation}
M^H= 2\pi r_+(1-\frac{r_-}{r_+})\, .
\end{equation}
This quantity always vanishes when $r_-=r_+$.
In this sense the black hole disappears completely in the
extremal limit.
The red-shifted temperature $T^H$ measured by a distant observer is
formally given by
\begin{equation}
T^H=\frac{\hbar}{2\pi k_B}|\kappa |= \frac{\hbar}{4\pi
k_Br_+}(1-\frac{r_-}{r_+})^{\frac{1-a^2}{1+a^2}}
=\frac{\hbar}{4\pi k_Br_+}(\frac{M^H}{2\pi r_+})^{\frac{1-a^2}{1+a^2}}
\, .
\end{equation}
{}From this it is clear that $T^H$ vanishes when $a<1$, it is finite when $a=1$
and
infinite whenever $a>1$ in the extremal limit.
By simply using the event-horizon area of the black hole as
a measure of the entropy content $S^H$ of the black hole we get
\begin{equation}
S^H=\frac{1}{4}A^H=\pi (r_+(\frac{M^H}{2\pi r_+})^{\frac{a^2}{1+a^2}})^2=
\frac{\hbar}{8\pi k_B}\frac{M^H}{T^H}\, .
\end{equation}
It follows that the entropy vanishes in the extremal state
independent of the value of $a$. It has been argued that
the thermo-dynamic interpretation breaks down near the extremal
limit \cite{Preskill,Holzhey}. Nevertheless we feel that the above relations
give an indication of the behaviour of the thermodynamical entities
in this limit, but we empasize the
vanishing of $M^H$.\\
\smallskip
The qualitative thermo-dynamical properties of the dilaton holes
in the extremal limit depend crucially on the value of the coupling
constant $a$. It is interesting to investigate whether this is
reflected in the dynamics of particles propagating
 in these geometries. Since
the case of a
real massless scalar field is treated in \cite{Holzhey}
we shall consider a classical photon and
the propagation of neutrinos.
Consider a photon
propagating in a general dilaton black hole background
 with fixed $\theta$ coordinate.  Define $E=p_t$ and $L=p_\phi$.
Then $g_{\mu\nu}p^\mu p^\nu =0$ implies
\begin{equation}
\dot{r}^2=E^2-\frac{L^2}{r^2}(1-\frac{r_+}{r})(1-\frac{r_-}{r})^{\frac{1-3a^2}{1+a^2}}\equiv E^2-V^2\, .
\end{equation}
{}From this it follows that
\begin{eqnarray}
\ddot{r}=-\frac{1}{2}\frac{dV^2}{dr}
=\frac{L^2}{r^3}(1-\frac{r_-}{r})^{\frac{1-3a^2}{1+a^2}}(1-\frac{3r_+}{2r}-\frac{r_-}{2r}(\frac{1-3a^2}{1+a^2})(\frac{r-r_+}{r-r_-}))\, .
\end{eqnarray}
In the extremal limit the analysis of this expression is particulary simple.
Due to the
exponent appearing in this expression it is natural to concentrate the
attention
to three $a$-parameter intervals: (I) $0\leq a< 1/\sqrt{3}$, (II)
$a=1/\sqrt{3}$
and (III) $a> 1/\sqrt{3}$. The second derivative
$\ddot{r}$ may vanish outside $r_+$ when $a$ takes values
in the parameter intervals (I) and (II) while turning points only exist outside
$r_+$ in region (III)
if we have the additional
restriction $a< 1$. When $a=1$ the turning point coincides
with $r_+$ and when $a>1$ the turning point is behind $r_+$.
Hence, in the extremal limit closed circular photon orbits can only exist
outside
$r=r_+$ when $a<1$ .
Another interesting question is whether $\ddot{r}$
may diverge. In (I) and (II) this function is finite for all non-zero values of
$r$.
When $1/\sqrt{3}< a< 1$ $\ddot{r}$ diverges on $r_+$. With $a=1$ the radial
acceleration
is finite on $r_+$ while for $a> 1$ it diverges again for
this value of the radial coordinate.
The significance of the $a=1/\sqrt{3}$ solution is somewhat
unclear. It does not seem to have any remarkable
 thermo-dynamical properties in the extremal limit. However, it is worth noting
that only $a=1$ and not
 $a=1/\sqrt{3}$ appears as a special value for the
scalar field dynamics in the extremal limit \cite{Holzhey}.\\
\smallskip

Let us now turn to the scattering of particles with
spin. In flat
space fermions can be described by two, in general coupled, equations
of two $2$-component spinors $P^A$ and $\overline{Q}_{\dot{B}}$. Following
\cite{Chan}
we write $P^0\equiv F_1$, $P^1\equiv F_2$, $\overline{Q}^{\dot{1}}\equiv G_1$
and
$\overline{Q}^{\dot{0}}\equiv -G_2$. In the Newman-Penrose formalism
the equations governing the dynamics of
massless fermions can be written as
\begin{eqnarray}
(D+\epsilon -\rho )F_1+(\delta^*+\pi -\alpha )F_2&=& 0\\
(\Delta +\mu -\gamma )F_2 +(\delta +\beta -\tau )F_1&=& 0\\
(D+\epsilon^*-\rho^*)G_2-(\delta +\pi^*-\alpha^*)G_1&=& 0\\
(\Delta +\mu^*-\gamma^*)G_1-(\delta^*+\beta^*-\tau^*)G_2&=& 0\, .
\end{eqnarray}
Using the standard null-frame notation $(\vec{l},\vec{n},\vec{m},\vec{m^*})$
it follows that $D=\vec{l}$ ,$\Delta =\vec{n}$, $\delta =\vec{m}$ and
$\delta^*=\vec{m^*}$. It is also assumed that
$l^an_a=-m^am_a^*=-1$ and $l^am_a=n^am_a=0$.
We will let $\vec{l}$ represent the four-velocity of an
outwards radially moving massless and electrically uncharged
particle with vanishing orbital angular momentum ($L=0$)
\begin{equation}
l^\mu = \frac{\sigma^2}{\Delta}\delta^a_0+\delta^a _1\, .
\end{equation}
Further, $n^\mu$ is identified with the four-velocity of the corresponding
inwards moving particle
\begin{equation}
n^\mu = \frac{1}{2}\delta^a_0-\frac{\Delta}{2\sigma^2}\delta^a_1\, .
\end{equation}
We also define
the complex vector $m^\mu$ ;
\begin{equation}
m^\mu = \frac{1}{\sqrt{2}r\sigma}(\delta^a_2+\frac{i}{\sin\theta}\delta^a_3)\,
{}.
\end{equation}
The non-zero spin coefficients are then given by
\begin{eqnarray}
\rho &=&\frac{-1}{r\sigma}\partial_r (r\sigma)\, ,\, \mu
=\frac{-\Delta}{2r\sigma^3}\partial_r(r\sigma )\\
\gamma &=&\frac{1}{4}\partial_r (\frac{\Delta}{\sigma^2})\, ,\,
\alpha =-\beta =\frac{-\cot\theta}{2\sqrt{2}r\sigma}\, .
\end{eqnarray}
Let $f_1=r\sigma F_1$, $f_2=F_2$, $g_1=G_1$ and $g_2=r\sigma G_2$.
Also define
\begin{eqnarray}
\hat{L}&=&\partial_\theta +\frac{i}{\sin\theta}\partial_\phi
+\frac{1}{2}\cot\theta\\
\hat{V}&=& i\omega\frac{\sigma^2}{\Delta}+\partial_r\\
\hat{W} &=&\frac{i}{2}\omega -\frac{\Delta}{2\sigma^2}\partial_r-
\frac{\Delta}{2r\sigma^3}\partial_r(r\sigma
)-\frac{1}{4}\partial_r(\frac{\Delta}{\sigma^2})\, .
\end{eqnarray}
We concentrate our attention to the F-equations since the G-equations are just
the
complex conjugate of these. Then eq.(25) becomes
\begin{equation}
\hat{V}f_1+2^{-1/2}\hat{L}^\dagger f_2=0
\end{equation}
and eq.(26) can similarly be written
\begin{equation}
r^2\sigma^2\hat{W}f_2+2^{-1/2}\hat{L}f_1=0\, .
\end{equation}
By utilizing that
\begin{equation}
\hat{W}=-\frac{\sqrt{\Delta}}{2r\sigma^2}\hat{V}^\dagger (r\sqrt{\Delta})
\end{equation}
eq.(38) can be brought to the more useful form
\begin{equation}
r\sqrt{\Delta}\hat{V}^\dagger (r\sqrt{\Delta}f_2)-2^{1/2}\hat{L}f_1=0\, .
\end{equation}
Define
\begin{eqnarray}
f_1&=&R_{1/2}(r)S_{1/2}(\theta )e^{i\omega t}e^{im\phi}\,\, ,\,\,
g_1=R_{-1/2}(r)S_{1/2}(\theta )e^{i\omega t}e^{im\phi}\\
f_2&=&R_{-1/2}(r)S_{-1/2}(\theta )e^{i\omega t}e^{im\phi}\,\, ,\,\,
g_2=R_{1/2}(r)S_{-1/2}(\theta )e^{i\omega t}e^{im\phi}\, .
\end{eqnarray}
where $m$ takes on positive and negative integer values.
Then the two equations above separate into
\begin{eqnarray}
\hat{V}R_{1/2}-\lambda R_{-1/2}&=&0\\
r^2\sigma^2\hat{W}R_{-1/2}+ \lambda R_{1/2}&=&0
\end{eqnarray}
and
\begin{eqnarray}
\hat{L}^\dagger S_{-1/2}+\sqrt{2}\lambda S_{1/2}&=&0\\
\hat{L}S_{+1/2} -\sqrt{2}\lambda S_{-1/2}&=&0\,
\end{eqnarray}
where $\lambda$ is the separation constant.
Define $P_{-1/2}=\sqrt{2}r\sqrt{\Delta}R_{-1/2}$ and $P_{1/2}=R_{1/2}$
and let $\sqrt{2}\lambda\rightarrow\lambda$. The radial
equations then become
\begin{eqnarray}
r\sqrt{\Delta}\hat{V}P_{1/2}&=&\lambda P_{-1/2}\\
r\sqrt{\Delta}\hat{V}^\dagger P_{-1/2}&=&\lambda P_{1/2}\, .
\end{eqnarray}
Further
define two functions $Z_\pm$ by $Z_\pm =P_{1/2}\pm P_{-1/2}$. Then
a one dimensional wave equation in $r^*$ follows
\begin{equation}
(\frac{d^2}{dr^{*2}}+\omega^2)Z_\pm =V_\pm Z_\pm\, ,
\end{equation}
where
\begin{equation}
V_\pm =\lambda^2\frac{\Delta}{r^2\sigma^4}\pm\lambda\frac{d}{dr^*}(
\frac{\sqrt{\Delta}}{r\sigma^2})\, .
\end{equation}
Solving this equation for $Z_{\pm}$ we can easily obtain the solutions
for $P_{\pm 1/2}$. When $a=0$ and $Q=0$ this expression reduces to
the one found in the Schwarzschild geometry. According to standard scattering
theory $V_-$ and $V_+$ will give rise to the same reflection and
transmission coefficients \cite{Chan}. It is easily seen that $V_\pm$ vanishes
on $r=r_+$ in the
extremal limit when $a<1$ and is finite outside.
When $a=1$ it is finite on $r=r_+$. When $a>1$ the potential is
singular at the horizon in the extremal limit. This is exactly what
 happens in the scalar field case. In addition to
this singularity structure the potential above exhibits an additional pole at
$r=r_+$ for
$a=\sqrt{3}$. This pole did not appear neither in the point particle case
nor in the dynamics of the real scalar field.
It is interesting to note that the value $a=\sqrt{3}$ arises in Kaluza-Klein
dilaton black hole solutions \cite{Frolov,Gibbons2,Duff}
and string solutions \cite{Duff}.
The equations for the $S_{\pm 1/2}$ functions are independent of $a$.
It follows that these functions are identical to the corresponding functions
in the Schwartzschild geometry, i.e. spherical harmonics $Y_{lm}(\theta ,\phi
)$.
The regularity of these functions on the unit sphere implies that $\lambda
=l(l+1)$.
Hence, for the lowest angular momentum mode $l=0$ it follows that
$V_\pm$ vanishes. This feature is also present in the massless
point particle case but
it is absent in the dynamics of a real massless scalar field. An immediate
consequence of the vanishing of $V_\pm$ is that the outgoing Hawking radiation
for $a>1$ holes in the extremal state apparently gives rise to a divergent
change of the gravitational mass $M^H$ relative to infinity
\begin{equation}
\frac{dM^H}{dt}|_{l=0}=\int\frac{\Gamma (l=0,\omega )\omega d\omega}{e^{\omega
/T^{H}}+1}
\rightarrow\infty\, .
\end{equation}
Since $V_\pm =0$ this implies that the transmission coefficient $\Gamma
(l=0,\omega )$
equals unity in the extremal limit and
seems to indicate a catastrophic deradiation
of these configurations. We interpret the result to mean
that we are outside the domain of validity of the
semiclassical approximation and that we are no longer free
to ignore the back reaction on the geometry. However, it would
seem that we could retreat sufficiently far away from
extremality for the thermodynamical results to be valid
and still obtain a very large rate for the outgoing Hawking radiation.
Naively this would seem to eventually result in a negative
gravitational mass of the black hole when $M^H$ is sufficiently
small.{\footnote{Note that although the potential (50) vanishes also for a
Reissner-Nordstr\"{o}m black hole there is no catastrophic mass-loss in that
case. This is because the temperature of those holes vanishes in the extremal
limit.}}
\\
\smallskip
\section*{Acknowledgements}
BJ thanks the University of Stockholm for hospitality during
the time which parts of this work was carried out.
UL acknowledges partial support from the NFR under
Grant No.F-FU 4038-300 and from NorfA under grant No. 93.35.088/00.


\end{document}